\documentclass[aip,amsmath,amssymb,preprint]{revtex4-2}
\usepackage{graphicx}
\usepackage{dcolumn}
\usepackage{bm}
\usepackage[usenames,dvipsnames]{color} 

\usepackage[utf8]{inputenc}
\usepackage[T1]{fontenc}
\usepackage{mathptmx}
\usepackage{etoolbox}
\usepackage{ulem}
\usepackage{xcolor}
\usepackage{natbib}
\usepackage{float}
\usepackage{subfigure}
\usepackage{url}       

\makeatletter
\def\@email#1#2{%
 \endgroup
 \patchcmd{\titleblock@produce}
  {\frontmatter@RRAPformat}
  {\frontmatter@RRAPformat{\produce@RRAP{*#1\href{mailto:#2}{#2}}}\frontmatter@RRAPformat}
  {}{}
}%
\makeatother

\begin{document}

\title[Assessment of the Runaway Electrons induced damage to the Tokamak First Wall]{Assessment of the Runaway Electrons induced damage to the Tokamak First Wall}
\author{L. Singh$^{1,2}$, M. De Bastiani$^2$, R. Bonifetto$^2$, F. Subba$^2$, D. Borgogno$^1$} 

\address{
Istituto dei Sistemi Complessi-CNR and Dipartimento di Energia, Politecnico di Torino, Torino 10129, Italy}
\address{
NEMO Group, Dipartimento Energia, Politecnico di Torino, Torino 10129, Italy}

\email{lovepreet.singh@polito.it}

\begin{abstract}
The study assessed the damage caused by Runaway Electrons (RE) on First Wall tiles, comparing the effects on Beryllium and Tungsten. This was done by using realistic RE energy distribution functions to replicate RE impacts through the FLUKA code. These energy distribution functions are based on the ASDEX Upgrade experiment \# 39012. The parametric analysis carried out with FLUKA in the presence of magnetic fields indicated a clear relationship between the beam impact angle and the material deposited energy, demonstrating that higher impact angles lead to deeper electron penetration and greater deposited energies. A finite element model based on apparent heat capacity formulation in FreeFem++ was developed to analyze the material thermal response to such thermal loads using volumetric energy density profiles from FLUKA simulations as input. Different RE current values were simulated to show its influence on the evolution of the material temperature and melting thickness.
    
\end{abstract}

\maketitle
%
\vspace{2pc}
\\
%
%
%

\section{\label{sec:intro}Introduction}

The impact of runaway electrons (RE) on Plasma Facing Components (PFC) in future Tokamak reactors, including ITER  \cite{Bandaru2025} and DEMO\cite{Vannini2025}, is a topic of significant concern. The potential damage caused by disruptions-induced RE presents a notable operational challenge for upcoming fusion machines. These electrons can deposit substantial energy within PFC, which could compromise their structural integrity and lead to melting \cite{Bartels1993}. Various studies have conducted numerical simulations to assess the damage caused by RE on PFC \cite{Igitkhanov2011b, Caloud2024, Ratynskaia2025}.

In Ref. \cite{Igitkhanov2011b}, the authors have examined the energy loss rate of runaway RE in ITER plasma-facing materials, namely Beryllium (Be) and Tungsten (W). In particular, the study has used the Monte Carlo Energy Deposition code  ENDEP \cite{Bazylev2011} and the fluid Melt Motion on Surface code MEMOS \cite{Bazylev2007} to evaluate energy deposition profiles and material erosion. For the energy deposition study (with ENDEP), the simulations have been performed assuming a Gaussian distribution of RE in momentum and different impact angles to consider the effect of the magnetic field on the RE incidence angle. With 1° impact angle only half of RE energy has been shown to be absorbed in Be while in W, the fraction of absorbed energy is 30\%. The rest of the energy is reflected off by back-scattered electrons. With 20° impact angle, the ratio of absorbed energy has been demonstrated to reach 80\% in Be and 50\% in W. The study does not consider the presence of a magnetic field, which causes a higher fraction to be deposited in the material. The study done with the MEMOS code considering a RE beam loss time of 10 ms and an energy of  25 $MJ/m^2$ predicted substantial erosion on the ITER FW Be bulk armor through calculations of RE energy deposition. Meanwhile, for W, it has been shown that the RE impact does not melt it.

In Ref. \cite{Caloud2024} the Monte Carlo code FLUKA \cite{Ahdida2022, Battistoni2015} and the finite element method (FEM) solver COMSOL Multiphysics \cite{COMSOL} has been used to design a calorimetry probe to measure the impact of a RE beam following a disruption. For the FLUKA simulations, a monoenergetic pencil (point) RE beam with a fixed impact angle has been used. The simulation from these codes has been used to define the requirements for sensor parameters, including temperature range, size, type, position, and the time resolution of the measured data. 

In Ref. \cite{Ratynskaia2025}, which represents the state of the art in the field, a workflow composed of different codes is proposed and applied to the study of the thermo-mechanical response of a graphite sample to controlled RE impact in DIII-D reactor. In particular, the Kinetic Orbit Runaway electrons Code KORC \cite{Carbajal2017} is used to determine the RE striking positions and momenta, the Monte Carlo transport code Geant4 \cite{Allison2016} is used to obtain the volumetric energy deposition and COMSOL for the thermoelastic response. The KORC output, regarding the energy and impact angle of the monoenergetic RE beam, which best matched the data from the eDIII-D experiment, is used as input for the volumetric energy deposition calculation with GEANT4. In GEANT4, the magnetic field is considered inside and outside the material to account for the backscattered electrons re-deposition. The modelling of the sample thermo-mechanical response to RE incidence is investigated the initiation of brittle failure.

In our study, we have established a workflow that employs FLUKA (version 4-3.1) and FreeFem++ \cite{MR3043640} to evaluate the damage induced by a RE beam onto Tokamak First Wall (FW) materials. In this paper, this workflow was applied to study the RE beam impact on ASDEX Upgrade (AUG) Tungsten (W) tile. In addition, the response of the W tile was compared to that of a Beryllium (Be) tile. The RE beam features were derived from the AUG experiment \#39012. Therefore, realistic RE energy distribution functions were used to simulate the impact. The magnetic field, an approximation of the AUG magnetic field at the wall, was defined inside and outside the material to account for the backscattered electrons re-deposition. To the best of our knowledge, as underlined above, the results present in the literature have focused on either a monoenergetic beam or analytical energy distribution functions.

The VED information obtained from FLUKA served as input for our newly developed thermal model in FreeFem++, which incorporates phase change physics. Using finite element techniques and an apparent heat capacity formulation, the model accounts for material melting \cite{ComsolDOC, richiusa2023advances}. Compared to the models used in previous studies, this approach is significantly simplified, focusing solely on the melting process while disregarding complex dynamics, such as those discussed in \cite{Ratynskaia2025}. However, since the primary objective of this work is to examine the onset of melting, this level of simplification is deemed sufficient.

The tool was extensively automated using a specialized Python wrapper, which manages input generation and simulation launching. This enhancement allows the tool to simulate various geometries and loads relevant to FLUKA analysis, increasing its versatility.

The thermal tool was benchmarked against results from the literature to validate its accuracy and reliability. Its application to a representative example demonstrated its efficacy in simulating transient phenomena.

Detailed information about the physics models used and implemented in the numerical tools can be found in Section \ref{sec:physics}. Additionally, the setup of simulations in FLUKA and FreeFem++ is outlined in Section \ref{sec:setup}. At the same time, the results and subsequent discussion are detailed in Sections \ref{sec:results} and \ref{sec:discussion}, respectively. Conclusions close the paper.

\section{\label{sec:physics} Summary of the physics models}

The following section briefly overviews the physics models employed in FLUKA and FreeFem++. Specifically, subsection \ref{subsec:FLUKAphysics} outlines the models utilized in FLUKA, while subsection \ref{subsec:thermalphysics} delineates the implementation of the apparent heat capacity method in FreeFem++ and the corresponding validation process.

\subsection{\label{subsec:FLUKAphysics} FLUKA physics models}

FLUKA is a versatile Monte Carlo code extensively used to analyse particle transport and their interactions with materials. Notably, the code can simulate a wide range of particles, including electrons, photons, and neutrons, making it suitable for various applications.

The FLUKA simulation software utilizes a multiple-scattering algorithm to model electron transport in materials, drawing on the Moliere thy for its foundation \cite{Ferrari1992}. This algorithm conserves CPU time by enabling multiple small-angle scatterings during each step without tracking the particle trajectory after each scattering event. The Moliere theory, which assumes small-angle deflections of charged particles within the material, derives a probability distribution for the particle scattering angle, considering the cumulative effect of numerous small deflections occurring during a step \cite{Ferrari1992}. Following each step, the code randomly samples the scattering angle from this probability distribution based on material and particle properties, with the energy lost during the step being distributed over the entire step. Additionally, the production of secondary electrons occurs if the energy lost during the step exceeds the threshold set for delta ray production \cite{Ferrari1992}.

In FLUKA, the selection of the optimal step length for sampling a particle mean free path is performed automatically based on material and particle properties using the multiple scattering method \cite{Ferrari1992}. This method ensures that the results retain insensitivity to the step size while enabling enhanced accuracy by imposing a maximum fraction of total energy to be lost in a step. This constraint governs the selection of the step size. The simulations conducted in our study utilize the default value for the maximum fraction of the total energy lost in a step, which is set at $20\%$ for electrons \cite{Ferrari1992}. Nevertheless, we have conducted tests to evaluate the appropriateness of the default value in the context of our studies, as discussed in section\ref{subsec:EnDep_fixed}.

\begin{figure}[!htbp]
\centering
\includegraphics[scale=0.4]{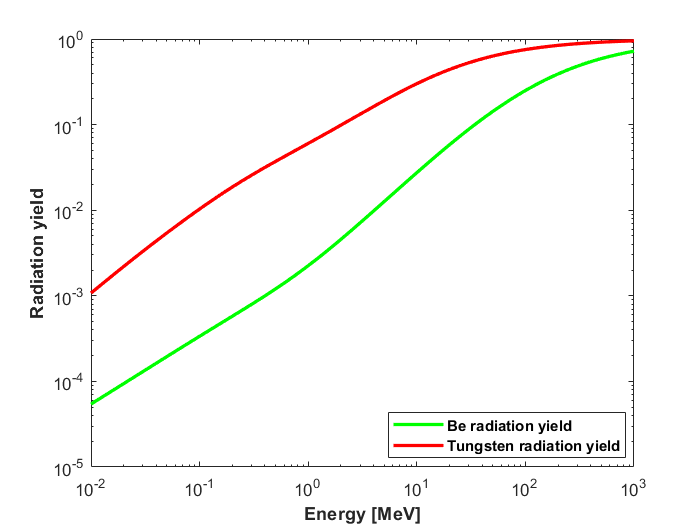}%
\caption{\label{fig:yield} Radiation yield as a function of electron energy in MeV for Beryllium (green) and Tungsten (red).}
\end{figure}

The cross-section for collision-stopping power, which represents the mean energy loss per unit path length due to ionization and excitation and scales as Z, is described by Moller and Bhabha scattering theories \cite{Fer05}. Conversely, the cross-section for radiative stopping power, denoting the average energy loss per unit path length due to Bremsstrahlung radiation and scaling as $Z^2$, is provided by the Berger and Seltzer (NIST) database \cite{Kim86}. Among the available energy distribution functions, we selected the beam with the highest maximum energy of approximately 7 MeV to simulate the worst-case scenario. At this energy, the collision-stopping power dominates over the radiative stopping power for both materials \cite{NIST}. However, in the case of Tungsten, radiation loss via Bremsstrahlung is comparable to ionization and excitation loss, resulting in significant photon production, as depicted in Fig. \ref{fig:yield}, showing the radiation yield as a function of electron energy for Tungsten (red) and Beryllium (green) \cite{NIST}.

The radiation yield, within the energy range of interest ($E_{kin}< 7 {MeV}$), denotes the mean fraction of the initial electron energy transformed into Bremsstrahlung energy as the electron decelerates to a complete stop, and it is significantly higher for Tungsten compared to Beryllium. The radiation length, which represents the distance at which particle energy decreases by a factor of 1/e, is higher for photons than electrons due to the different nature of the transport mechanisms of the two particles. This allow photons to penetrate deeply into the material and deposit their energy in regions far from the impact surface.

FLUKA is a highly adaptable tool that enables the study of electron and photon transport in materials in various manners. In the present analysis, the influence of Brehmstrahlung on the overall energy deposition pattern occurring during RE-material interaction has been investigated by adjusting the photon production threshold. The minimum threshold energy for photon transport and production in FLUKA is 100 eV, while it is set to 1 keV for electrons. For our study, we moved the threshold on photon production to 1 GeV to examine far-from-surface volumetric energy deposition profiles in different materials.

In FLUKA, the continuous slowing-down approximation (CSDA) is employed to model the transport of particles with energies below the thresholds set for delta-ray production \cite{Fer05}. The CSDA calculates the distance that a particle can travel with a specific energy before reaching the threshold at which it is considered to deposit all its energy at that point. This distance, known as the CSDA range, is determined using the total stopping power of the material, which is the sum of collision and radiative stopping powers. The total stopping power is assumed to be equivalent to the rate of energy loss along the particle path \cite{NIST}.

\begin{figure}[!htbp]
\centering
\includegraphics[scale=0.4]{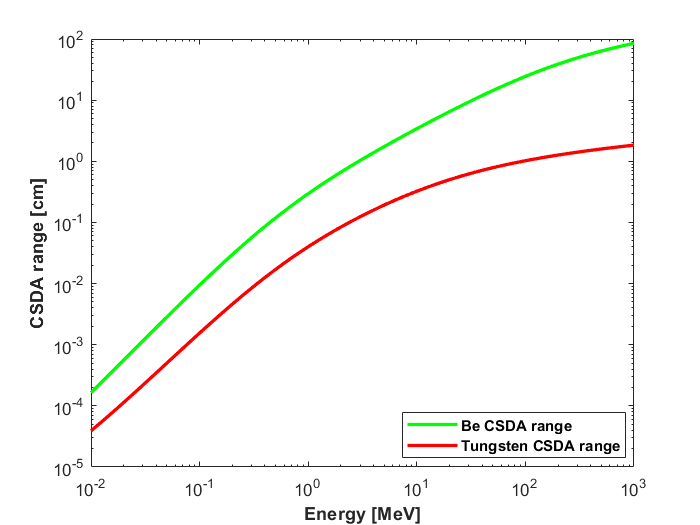}%
\caption{\label{fig:CSDA} CSDA range in cm as a function of energy in MeV for Be (green) and W (red).}
\end{figure}

To precisely simulate low-energy particle energy deposition, it is essential to appropriately tune the grid resolution along relevant directions and the threshold values set for particle absorption by the material. Fig. \ref{fig:CSDA} shows the CSDA range for electrons as a function of energy for Be (green) and W (red). The CSDA range for the different materials, sourced from \cite{NIST} (in $g/cm^2)$, was divided by the material density ($d_{Be}=1.8 g/cm^3$ and $d_W=19.25 g/cm^3)$ for comparison with the bin dimension in cm. A bin corresponds to the length of each cell into which the detector spatial grid is divided along the X, Y, and Z axes. 

If the bin dimension along the analyzed direction exceeds the CSDA range, the particle will stop at an incorrect location. For example, suppose a transport threshold of 100 keV is imposed. In that case, corresponding to a CSDA range of 1.6e-3 cm for Tungsten, the bin dimension along the Z direction should be comparable with 1.6e-3 cm to ensure accurate stopping depth.

\subsection{\label{subsec:thermalphysics} Thermal model implementation and validation}

Runaway electrons in the plasma-facing component can potentially heat the material to the point of melting, necessitating the use of a detailed thermal model that can accurately simulate the physics of the phase change to assess the expected PFC damage reliably.

Modelling the melting phenomena falls under the category of moving boundary problems as the position of the interface between the two phases is unknown \cite{zerroukat}. Over the past few decades, various techniques have been developed to address this type of problem, including variable time step grids  \cite{gupta1981variable}, integral methods \cite{caldwell2003nodal}, and boundary immobilization techniques \cite{caldwell2004numerical}. While these techniques are effective for 1D problems where a single point unequivocally determines the interface position, their applicability to 2D (or even 3D) cases, such as those examined in this study, may be limited due to the potentially complex shape of the two-phase interface. As a result, the apparent heat capacity formulation, previously utilized in Ref. \cite{richiusa2023advances}, has been employed in this study. This approach provides an implicit technique for determining the phase change interface by solving a single heat diffusion equation using an effective specific heat that accounts for the latent heat of fusion \cite{ComsolDOC}.

The apparent heat capacity method assumes that the phase change occurs within a temperature range centred around the melting temperature ($T_m$) with a width of$\delta$. In other words,  a smoothed phase transition in temperature around the actual melting temperature, characterized by the temperature "width" delta, is assumed. This allows to keep the problem differentiable from numerical point of view.During this range, the material is considered a mixture of the two phases, and the fraction of the initial phase is represented by a differentiable function $\theta$. The material density and its specific enthalpy can be expressed as a combination of the two phases.

\begin{equation}
    \rho = \theta \cdot \rho_{1} + (1-\theta)\cdot \rho_{2}
\end{equation}
\begin{equation}
    h = \frac{1}{\rho} \cdot (\theta  \rho_{1}  h_{1} + (1-\theta)  \rho_{2}  h_{2})
\end{equation}

Following the specific enthalpy definition, deducing the particular heat capacity expression is feasible, as denoted in equation \ref{cp}.

\begin{equation}
    c_p = \frac{\partial h}{\partial T} = \frac{1}{\rho}\cdot (\theta_1 \rho_1 c_{p,1} + (1-\theta)\rho_2 c_{p,2}) + \Delta h_f \cdot \frac{d\alpha_m}{dT}
    \label{cp}
\end{equation}

In the given equation, $\Delta h_f$ is the latent heat of fusion, the symbol $\alpha_m$ denotes the mass fraction, which is set to $-1/2$ before the occurrence of the phase change and switches to $1/2$ following the transformation. The precise definition of $\alpha_m$ can be found in equation \ref{am}.

\begin{equation}
    \alpha_m = \frac{1}{2} \cdot \frac{(1-\theta)\rho_2 - \theta \rho_1}{\rho}
    \label{am}
\end{equation}

The apparent heat capacity method implemented in this work utilizes FreeFEM++ \cite{MR3043640}, an open-source Finite Element software. The FreeFEM++ solver has been integrated into a Python wrapper, automatically generating the necessary inputs and initiating the execution process. This implementation enhances the tool adaptability, allowing it to potentially be used for any geometry and load configurations in FLUKA analysis.

The hyperbolic tangent function, denoted by $\theta$, has been selected due to its differentiability and possession of two horizontal asymptotes. The specific form of the chosen $\theta$ function is provided in equation \ref{theta}, with $\delta = 10.0 \, K$ being the selected parameter value.

\begin{equation}
    \theta (T) = \frac{1}{2} \cdot (1+tanh(T_m -T))
    \label{theta}
\end{equation}

The argument of $tanh$ is not divided by $\delta$ since the division would strongly enlarge the interval in which the phase concentration is strongly different from 0 or from 1, basically enlarging the "melting interval". Considering the hyperbolic tangent centered in the melting temperature and not deformed produces a continuous variation from (almost) 1 to (almost) 0 in the interval $\left (T_m-\frac{\delta}{2}, T_m+\frac{\delta}{2} \right )$. For the computation the derivative of the phase concentration is computed in the above mentioned interval, while it is set to 0 outside.

Assuming that the density of the material is constant through the phase change, obtaining a simplified expression for the mass fraction that is easy to derive is possible.
\begin{equation}
    \alpha_m = \frac{1}{2} - \theta(T)
\end{equation}

It is important to note that the expression for the latent heat contribution to the specific heat $(c_L)$, as denoted in equation \ref{cl}, can be obtained and is applicable within the temperature range associated with the phase change. It is pertinent to highlight that this contribution becomes null outside this temperature range.

\begin{equation}
    c_L(T) = \Delta h_f \cdot \frac{d\alpha_m}{dT} = \Delta h_f \cdot \frac{1}{2} \cdot \left( \frac{1}{cosh(T_m-T)} \right)^2
    \label{cl}
\end{equation}

A straightforward benchmark problem from \cite{zerroukat} has been utilized to validate the developed thermal tool and reported in the Appendix.

\section{\label{sec:setup} Simulation setups}

\begin{figure}[!htbp]
    \centering
    \includegraphics[scale = 0.8]{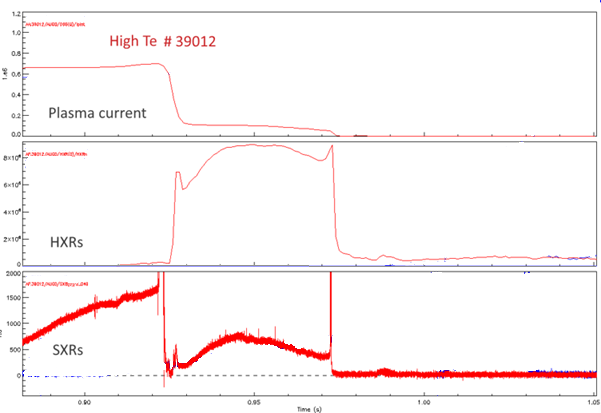}
    \caption{Plasma current, Soft X-ray and Hard X-ray evolution during the AUG experiment \# 39012. Adapted from \cite{Eurofusion_Wiki}.}
    \label{fig:experiment}
\end{figure}

The features of the RE beam used for simulating the impact through FLUKA and FreeFEM++ were derived from the beam observed during the ASDEX Upgrade (AUG) experiment \# 39012, whose dynamics in terms of the plasma current (top panel), Hard X-ray (HXR) (middle panel) and Soft X-ray (SXR) (bottom panel) is shown in Fig. \ref{fig:experiment}. As can be observed, around $t=0.93 s$ all the plasma current is carried by the RE current whose presence is confirmed by the HXR and SXR signals. After this time point a gradual decrease of the RE current amplitude takes place until around $t=0.97 s$ where the majority of the RE current is lost on a very short time. 

A FLUKA simulation requires the definition of a geometry, a source, a grid resolution, energy thresholds for electron and photon production transport and a magnetic field vector. Details about the required inputs are contained in subsection \ref{subsec:FLUKAinput}. 

The material properties needed for the assessment of the thermal damage, along with the other necessary parameters, are described in subsection \ref{subsec:FreeFeminput}.

\subsection{\label{subsec:FLUKAinput} FLUKA inputs definition}

The geometry pertains to a FW tile with measurements of 10x10 cm along the X (toroidal) and Y (poloidal) directions and a thickness of 3 cm along the Z (radial) direction. The entire runaway current is assumed to be lost to this single target plate.

\begin{figure}[!htbp]
    \centering
    \includegraphics[scale = 0.5]{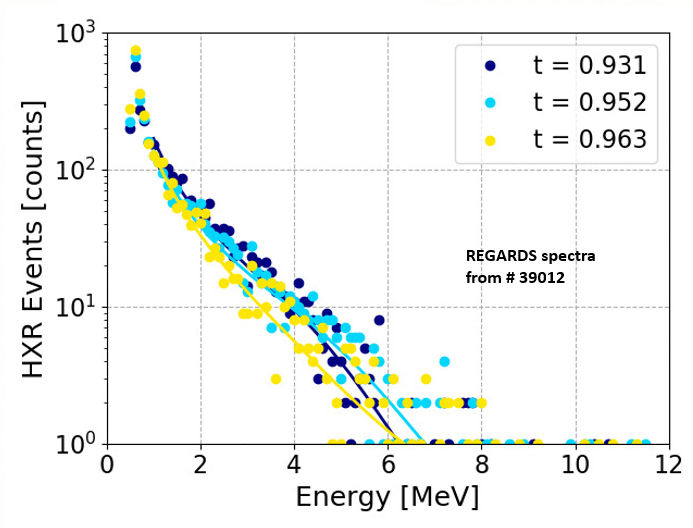}
    \caption{Plasma current, Soft X-ray and Hard X-ray evolution during the \#39012 AUG experiment. \cite{Eurofusion_Wiki}.}
    \label{fig:Hard_Xray}
\end{figure}

For the analysis of the RE impact through FLUKA, the RE energy distribution function was derived from the Hard X-ray data, shown in Fig. \ref{fig:Hard_Xray} at three different time steps. The Hard X-ray emission from RE has been measured through the Runaway Electron GAmma-Ray Detection System (REGARDS) spectrometer \cite{DalMolin2023}. To infer the RE distribution function from measured hard X-ray emission, first-order Tikhonov regularization has been used. The modelling of the  RE bremsstrahlung emission is based on the formula from Ref. \cite{Salvat2006}, and the detector response function has been computed using MCNP \cite{TechReport_2022_LANL_LA-UR-22-30006Rev.1_KuleszaAdamsEtAl}. Full details about the derivation can be found in \cite{DalMolin2020}. 

\begin{figure}[!htbp]
\centering
\includegraphics[scale=0.5]{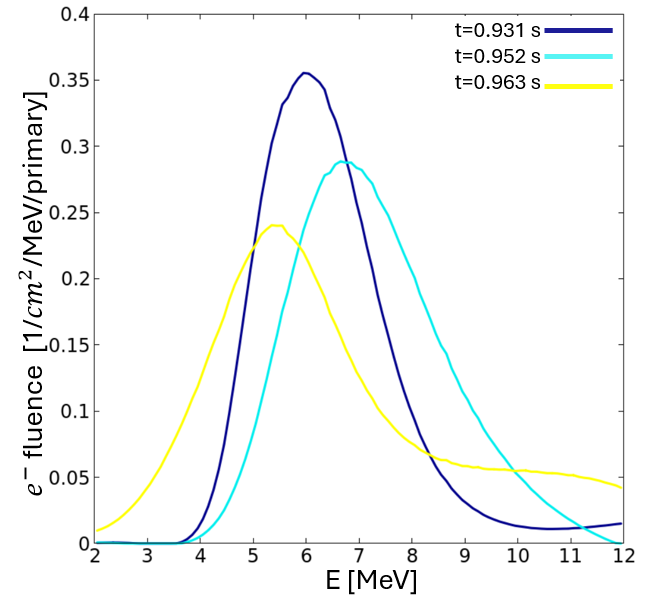}%
\caption{\label{fig:REdistribution} RE energy distribution in terms of the electron fluence ($1/cm^2/MeV/primary$) at t= 0.931s (blue), t=0.952s (cyan) and t=0.962s (yellow)}
\end{figure}

The derived RE energy distribution of the runaway electron functions are shown in Fig. \ref{fig:REdistribution} for the same time points of the Hard X-ray data. Among these energy distributions shown in terms of the electron fluence ($1/cm^2/MeV/primary$), the one corresponding to t=0.952 s (cyan) was selected to simulate the impact of a beam with the highest average electron energy $E_{kin}$=7.23 MeV. Therefore, it is assumed that the RE beam impact on the wall when $t=0.952$ s (instead of around 0.97 s).  

Two different RE spatial distributions were considered: a point beam and a 10 $cm^2$ square beam. In the case of a point beam, the particles originate from a single point (the impact location defined in (X,Y,Z) coordinates) in space (with no spatial extent), meaning all particles in the beam are emitted from the same position. It does not represent a realistic scenario. However, it is useful to analyze electron transport mechanisms within the material. On the other hand, the finite extension beam is utilized for thermal analysis.

The beam impacts the tile at the centre of its surface, located at coordinates (X,Y,Z) = (5,5,0). With a point beam, all the RE are concentrated in the impact position. A parametric scan is conducted to examine the effects of different impact angles within the range of (0,90) degrees on the ED profiles in Be and W. These impact angles are measured with respect to the tile surface along the toroidal direction.

Regarding domain resolution,the  tiles  were divided into a spatial grid with bin dimensions depending on the impact angle. For Be, with an angle above 5 degrees the resolution of (X,Y,Z) = (200, 200,60) $\mu m$ was adopted. Below this angle for Be and for every angle for W the grid resolution shown in Tab. \ref{tab:resolution} was adopted.

\begin{table}[!hbtp]
    \centering
    \begin{tabular}{|c|c|}
         \hline
        Z $[\mu m]$ & resolution $[\mu m]$ \\
         \hline
        <10 & (200, 200, 0.02) \\
         \hline
        <50 & (200,200, 0.08) \\
         \hline
        >50 & (200,200, 60) \\
        \hline
        
    \end{tabular}
    \caption{Grid resolution for at different radial points}
    \label{tab:resolution}
\end{table}

In the near surface resolution a bin dimension of the order of nanometers was adopted to accurately solve the steep gradients in the volumetric energy density profiles. Moreover, the grid resolution along the Z direction is lower than the CSDA range of 1e-2 cm, corresponding to the 100 keV electron transport threshold for Beryllium and  the CSDA range of 1.6e-3 cm for Tungsten.  

Regarding the magnetic field vector at the wall, approximated values were used. In particular, the magnetic field is expected to closely resemble the equilibrium field during the early stages of the collapse, meaning that there should be a minimal normal component. In addition, the relationship between the poloidal and toroidal components is determined by the value of the safety factor. Since RE beams are circular and typically carry lower current in high magnetic fields, it has been assumed that the ratio of toroidal to poloidal magnetic fields $B_{tor}/B{pol}=10$, and the ratio of toroidal to radial magnetic fields $(B_{tor}/B_{rad})=100$ \cite{Ondrej}. Considering the maximum magnetic field of AUG of 3.2 T at the magnetic axis, $R_0=1.6 m$ (major radius) and $a=0.8$ (minor radius), the magnetic field components at the wall are assumed to be $B_x=2.12 T$, $B_y=0.21 T$, and $B_z=0.02 T$. 

\subsection{\label{subsec:FreeFeminput} FreeFem++ inputs definition}

A $10 \,\, cm^2$ square, 5-degree inclined beam was considered in the examined cases. The cutting plane was set at the centre of the beam, parallel to the beam itself, in order to assess the slight inclination effect. To derive the power deposition distribution, the energy per unit volume, per primary computed by FLUKA, was multiplied by the number of involved particles (a function of the RE beam current) and then divided by the characteristic beam loss time. The loss time was fixed to 1 ms which corresponds to the duration of the HXR and SXR last spike observed in  Fig. \ref{fig:experiment}. The spike is given by the loss of the RE current to the wall. The resulting distribution was considered the sole driver of the problem, with adiabatic boundary conditions imposed on the tile sides. For the scope of this study, only the deposition duration was considered. The tile initial temperature, denoted as $T_0$, was fixed to 623 K.

Material properties, in terms of melting temperature ($T_m$), latent heat of fusion ($\Delta h_f$) and density ($\rho$), were obtained from \cite{tolias2017W} and \cite{tolias2022Be} for Tungsten and Beryllium, respectively. Table \ref{MatProp} presents the melting temperature and latent heat of fusion for both materials, as considered in the analysis.
\begin{table}[!htbp]
    \centering
    \begin{tabular}{|c|c|c|c|}
        \hline
        \textbf{Material} & \textbf{$T_{m}$ [K]} & \textbf{$\Delta h_f \,\, \left [ \frac{kJ}{kg} \right ]$} & \textbf{$\rho \,\, \left [ \frac{kg}{m^3} \right ]$}\\
        \hline
        Tungsten \cite{tolias2017W} & 3695.0 & 271.98 & 19250.0\\
        \hline
        Beryllium \cite{tolias2022Be} & 1560.0 & 1643.8 & 1800.0\\
        \hline
    \end{tabular}
    \caption{Melting temperature ($T_m$), latent heat of fusion ($\Delta h_f$) and density ($\rho$) of the considered materials adopted in the simulation.}
    \label{MatProp}
\end{table}

In the experiment, the RE current amplitude is approximately 100 kA at the beam loss time point, as can be observed in the top figure of Fig. \ref{fig:experiment} showing the evolution of the plasma current. However, a scan was carried out in this study to consider the damage caused by higher RE current amplitude. Therefore, the RE current was scanned within the 100 - 500 kA range while maintaining a fixed event duration of 1 ms. Moreover, all results will be presented with time normalized to the beam loss time.

\section{\label{sec:results} Results}

The volumetric energy density profiles with a fixed impact angle of 5 degrees are detailed in subsection \ref{subsec:EnDep_fixed}, where a point and a $10 \,\, cm^2$ beam are examined. Given the significant impact of the RE beam angle on the RE-material interaction, the outcomes of a parametric investigation concerning the energy deposition profiles are explained in subsection \ref{subsec:EnDep_scan}.

As for the thermal analysis, the assessment focuses on the $10 \,\, cm^2$ beam with a 5-degree impact angle, with results illustrating the material temperature and the evolution of melted thickness in \ref{subsec:thermal}.

\subsection{\label{subsec:EnDep_fixed} Volumetric energy density profiles with fixed impact angle}

\begin{figure}[!htbp]
\centering
\includegraphics[scale=0.4]{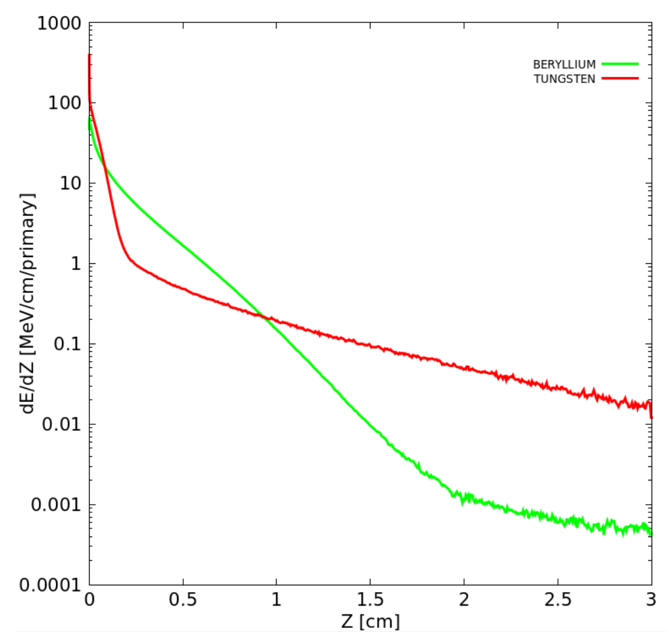} 
\caption{\label{fig:EnDep1st} 1D projection of the ED along the Z direction for Be (green) and W (red) with a point beam impacting at (X,Y,Z)=(5,5,0) cm with 5 degrees impact angle with respect to the tile surface }
\end{figure}

\begin{figure*}[!htbp]
\centering
\includegraphics[width=\textwidth]{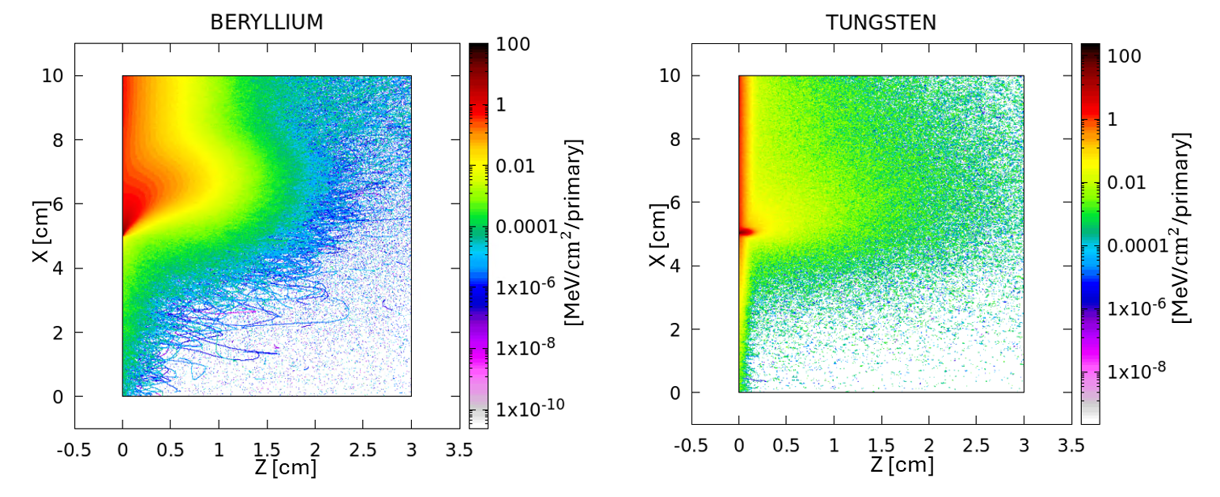}%
\caption{\label{fig:2D_BeW} 2D projection of the ED on the XZ plane for Be at the left panel and W at the right panel with a point beam impacting at (X,Y,Z)=(5,5,0) cm with 5 degrees impact angle with respect to the X direction.} 
\end{figure*}

The simulation setup utilized an impact angle of 5 degrees to compare the volumetric energy density (VED) in Beryllium and Tungsten. The 1D projection along the Z direction of the VED is depicted in Fig. \ref{fig:EnDep1st}, while the 2D projection of the VED on the XZ plane is presented in Fig. \ref{fig:2D_BeW}. The figure displays the ED for Be in the right panel and W in the left panel. The deposited energy is quantified in terms of $MeV/cm^3(primary$, computed by averaging over the third dimension Y.

A beam with an average energy of 7.23 MeV per electron deposits 5.17 MeV per beam electron in Be (green) and 5.23 MeV in W (red), as shown in Fig. \ref{fig:EnDep1st}, with an average statistical error below $3\%$. The remaining energy is carried by the back-scattered electrons and in a small fraction by the photons existing the computational domain. Notably, reducing the maximum fraction of energy to be lost in a single step to $5\%$ causes only a $2\%$ reduction in the deposited energy per electron, thereby allowing the default value of $20\%$ to be retained.

Figures \ref{fig:EnDep1st} and \ref{fig:2D_BeW} depict that the energy deposition resulting from beam electrons is notably more concentrated in W compared to Be, specifically within the first few millimetres of the material. It is observed that beam electrons undergo substantial attenuation in W compared to Be, leading to a two-order-of-magnitude reduction in energy within the first 2 mm of material depth. This behaviour can be attributed to W being a high Z material (Z=74) compared to Be (Z=4), making inelastic collisions more effective in W. Consequently, the runaway electron beam decelerates rapidly in W and loses its energy near the surface.

The comparison in Fig. \ref{fig:2D_BeW} reveals a notable difference in the behaviour of electron showers in Beryllium and Tungsten. The electron shower penetrates deeper into the material in Be before losing its initial directional information. In contrast, the shower acquires an isotropic angular distribution in W after traversing the first 2 mm. This disparity is attributed to the material elastic scattering dependency on the $Z^2$. Specifically, in W, elastic scattering is approximately two orders of magnitude more effective ($74^2/4^2$) compared to Be.

\begin{figure*}[!htbp]
\centering
\includegraphics[width=\textwidth]{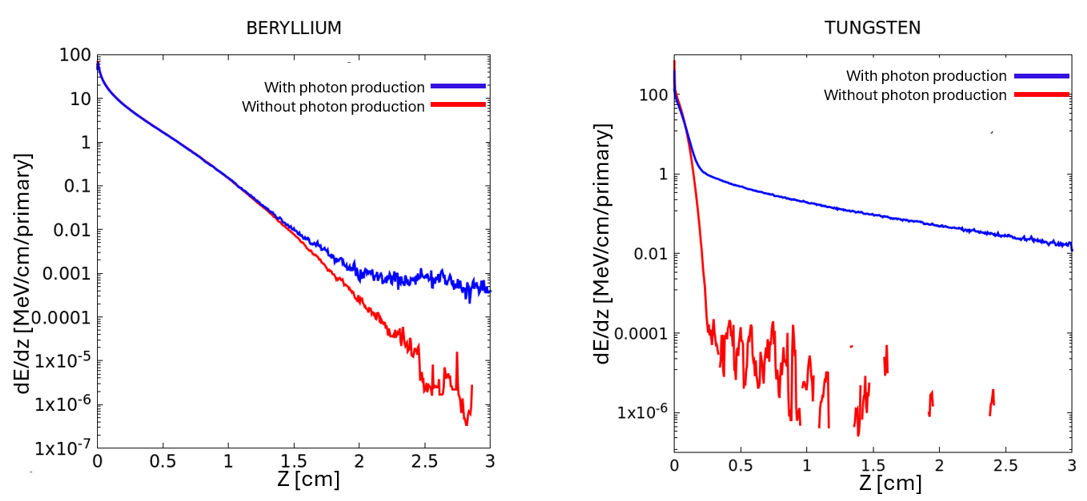}%
\caption{\label{fig:Photon_Nophoton} 1D projection of the ED in Be at the left panel and W at the right panel with (blue) and without (red) photon production.} 
\end{figure*}

With respect to electrons, photons become the primary energy carriers deeper in the material due to their high radiation length, as mentioned in the subsection \ref{subsec:FLUKAphysics}. Fig. \ref{fig:Photon_Nophoton} illustrates this through the 1D projection of the VED profile in Beryllium (left panel) and Tungsten (right panel) with default (blue) and high (red) photon production thresholds. In FLUKA, the default photon production threshold is set to 33 keV \cite{Fer05}, while a threshold of 1 GeV is imposed to explore the energy deposition pattern without photon production. In the case of Beryllium, photons primarily influence the energy deposition profile beyond 1.5 cm, where the deposited energies are less than 10 keV/cm. Conversely, in Tungsten, photons become the leading energy carriers after approximately 3 mm, where the deposited energy is still around 1 MeV/cm for a radiation source with an energy of roughly 7 MeV per primary electron.

\begin{figure*}[!htbp]
\centering
\includegraphics[width=\textwidth]{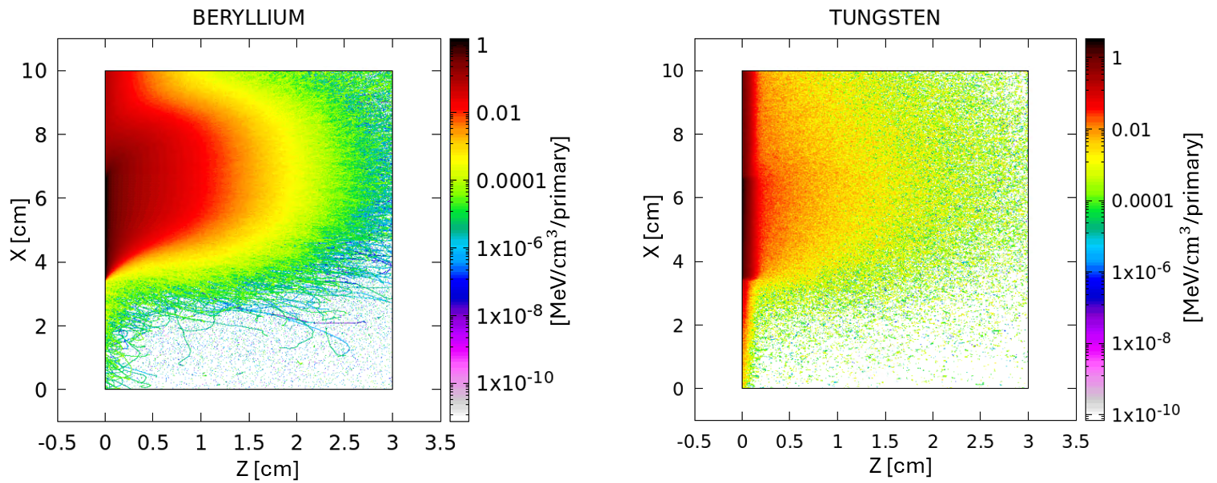}%
\caption{\label{fig:BeW10} 2D projection of the ED on the XZ plane for Be at the left panel and W at the right panel with a $10 cm^2$ beam impacting at (X,Y,Z)=(5,5,0) cm with 5 degrees impact angle with respect to the X direction.} 
\end{figure*}

The outcomes from considering a 10 $cm^2$ beam impact closely resemble a point beam. Specifically, the 1D projection of the VED and the amount of energy deposited per beam particle exhibit similar characteristics for both Beryllium and Tungsten in the two scenarios. The difference arises in the VED, with lower values observed in the case of a 10 $cm^2$ beam. Due to the a broader spatial distribution, the VED becomes less localized, as illustrated in Fig. \ref{fig:BeW10}.

\subsection{\label{subsec:EnDep_scan} Volumetric energy density profiles with different impact angles}

In Subsection \ref{subsec:EnDep_fixed}, a fixed impact angle was used to compare the VED profiles in different materials and to highlight the influence of various electron transport mechanisms on these profiles. However, in a plasma environment, electrons travel in a spiral trajectory along magnetic field lines, resulting in varying angles at which the electrons impact the material. Although FLUKA does not currently consider electron gyration, a thorough analysis can be performed to generate VED profiles at different impact angles, enabling a comprehensive examination of their effects on the deposition pattern. Specifically, the focus is on impact angles ranging from 0 to 90 degrees.

\begin{figure*}[!htbp]
\centering
\includegraphics[width=\textwidth]{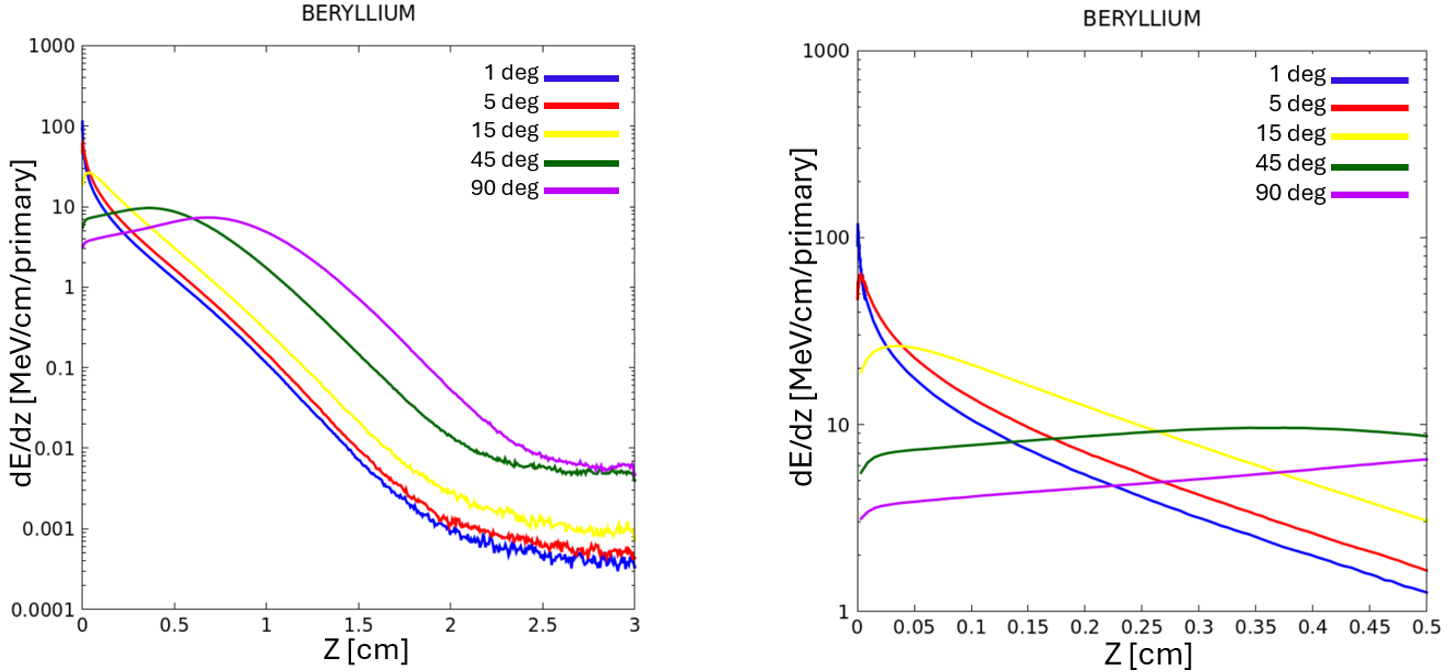}%
\caption{\label{fig:Be_impact} 1D projection of the ED in Be over the full thickness at the left panel and near the surface at the right panel.} 
\end{figure*}

\begin{figure*}[!htbp]
\centering
\includegraphics[width=\textwidth]{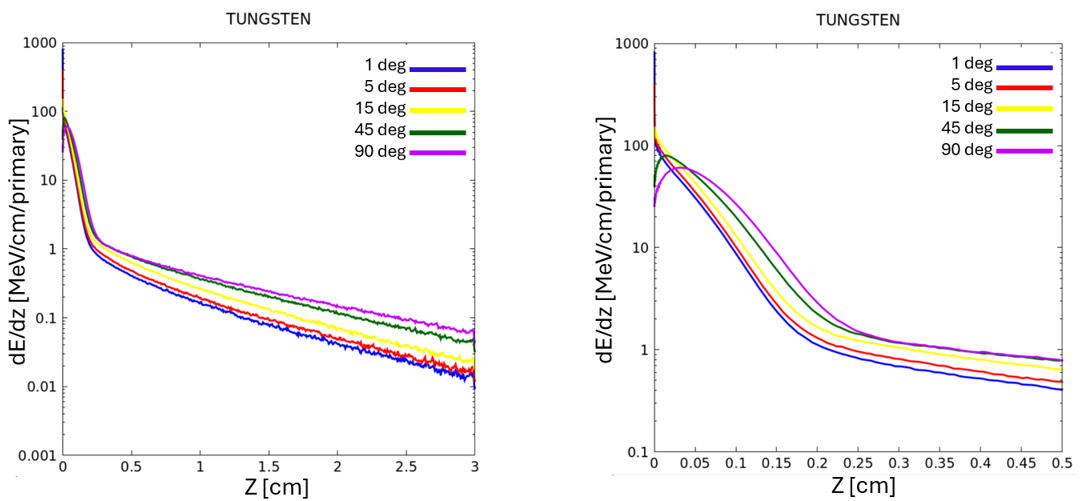}%
\caption{\label{fig:W_imapact} 1D projection of the ED in W over the full thickness at the left panel and near the surface at the right panel.} 
\end{figure*}

Figures \ref{fig:Be_impact} and \ref{fig:W_imapact}, depicting the VED profiles with different angles for Beryllium and Tungsten, illustrate a decrease in the  maximum of the volumetric energy density and a shift of the maximum deeper into the material.  At a 90° angle, the peak deposition shifts to 1 cm for Beryllium and approximately 0.5 mm for Tungsten.


The displacement of the VED maximum is given by the inclination of the region characterized by the highest VED which corresponds to the blob around the initial direction of the incident electron beam. At a grazing impact angle, like the case with a 5° impact angle in Be, the region is almost on the surface of the material, resulting in a monotonously decreasing ED profile, as depicted in Fig. \ref{fig:EnDep1st}. Conversely, at a 15° impact angle, the region is entirely within the material, causing the maximum to shift to greater depths. Furthermore, the decrease in the VED maximum is attributed to the inclination of the most intense region at a 15° angle, which has a broader extent compared to the case where it is nearly vertical with a 5° impact angle.

The average energy deposited in the material per primary increases with the impact angle due to a lower number of back-scattered electrons. The observed average energy  per primary deposited in the materials is shown in Tab. \ref{tab:energy_Be} for Be and Tab. \ref{tab:energy_W} for W.

\begin{table}[!hbtp]
    \centering
    \begin{tabular}{|c|c|}
         \hline
        Impact angle & E [MeV] \\
         \hline
        1 & 4.17 \\
         \hline
        5 & 5.17 \\
         \hline
        15 & 6.68 \\
        \hline
        45 & 7.09 \\
        \hline
        90 & 7.1 \\
        \hline
        
    \end{tabular}
    \caption{Average energy deposited in Beryllium per primary with increasing impact angle}
    \label{tab:energy_Be}
\end{table}

\begin{table}[!hbtp]
    \centering
    \begin{tabular}{|c|c|}
         \hline
        Impact angle & E [MeV] \\
         \hline
        1 & 4.58 \\
         \hline
        5 & 5.23 \\
         \hline
        15 & 6.30 \\
        \hline
        45 & 6.78 \\
        \hline
        90 & 6.84 \\
        \hline
        
    \end{tabular}
    \caption{Average energy deposited in Tungsten per primary with increasing impact angle}
    \label{tab:energy_W}
\end{table}

\begin{figure*}[!htbp]
\includegraphics[width=\textwidth]{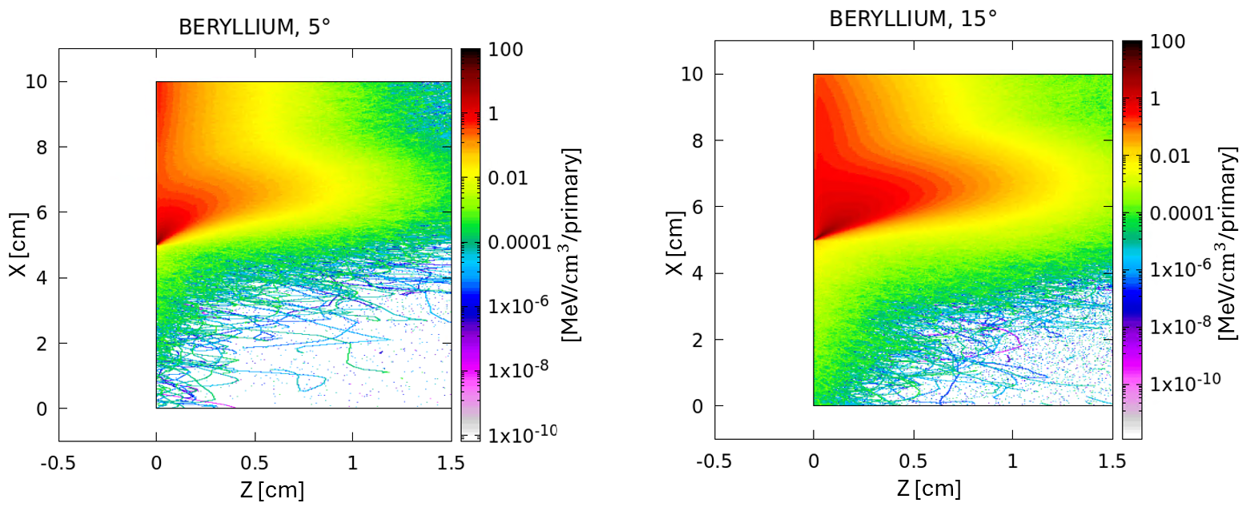}%
\caption{\label{fig:Be_ED} 2D projection of the ED on the XZ plane for Be with 5° at the left panel and with 15° at the right panel.} 
\end{figure*}

\subsection{\label{subsec:thermal} Thermal analysis of the RE beam power deposition}

Upon confirming the accuracy of the model in describing the relevant physics, it was utilized to simulate the thermal evolution of Tokamak FW tiles under the influence of RE beams. The energy deposition distributions obtained through the FLUKA code were input to the thermal model, with the 5-degree square beam case as the reference scenario.

Although the energy deposition distribution obtained with FLUKA is inherently 3D and depends on beam geometry and inclination, it has been demonstrated that a 2D model suffices to assess potential material melting. By comparing the RE beam loss time with the characteristic time of heat diffusion (given by the ratio between the square of the characteristic dimension and the material thermal diffusivity), it becomes apparent that heat diffusion significantly outweighs the beam loss time across the entire temperature range. Consequently, damage is expected near the beam contact area, where diffusion is secondary during the impact of the RE beam. 

In the case of a beam perpendicular to the tile, a 1D model along the beam axis is deemed adequate for estimating the melting thickness. A 2D model has been adopted to accommodate the analysis of inclined beam effects and the beam boundary region where diffusion-related effects may be macroscopically detected.

\begin{figure}[!htbp]
    \centering
    \subfigure[]{\includegraphics[scale=0.3]{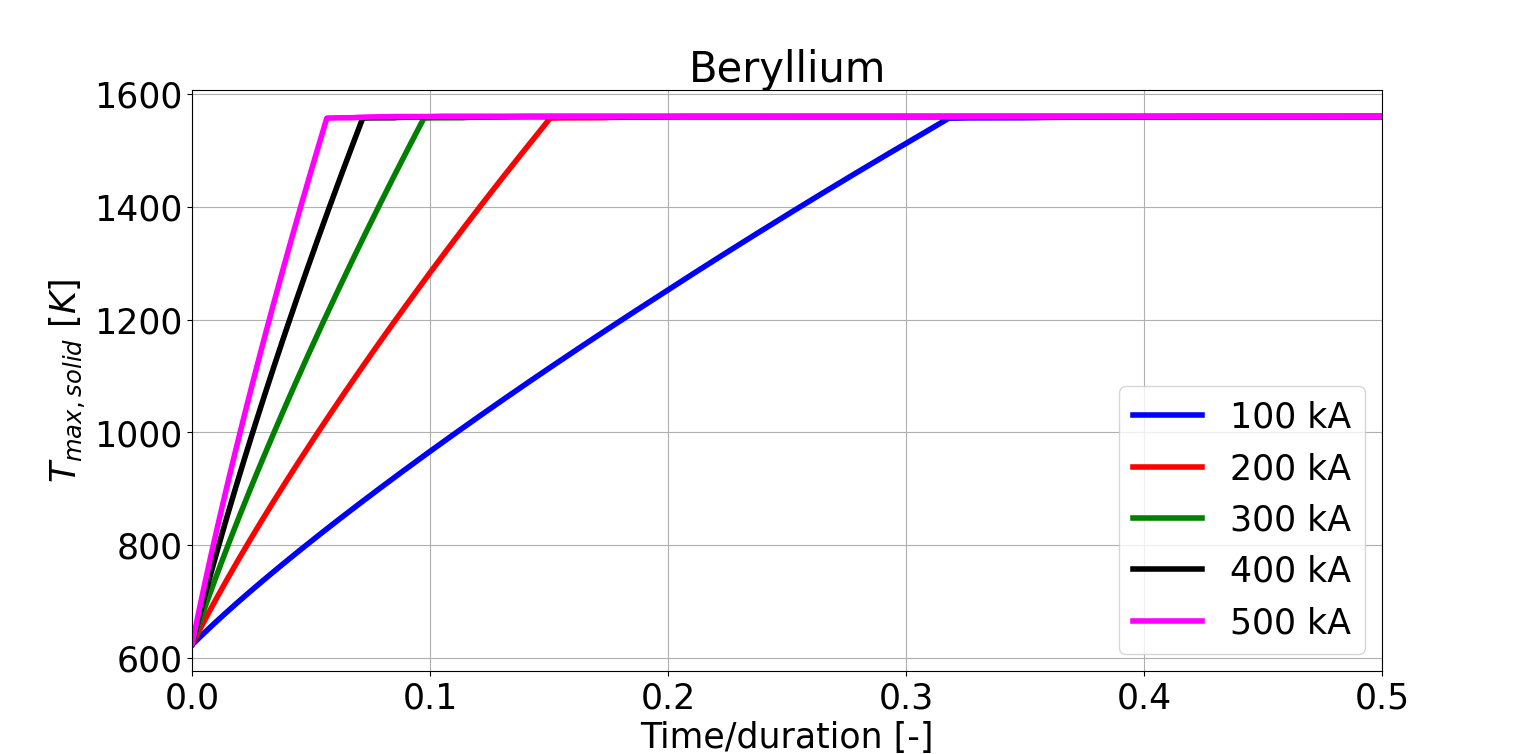}}
    \subfigure[]{\includegraphics[scale=0.3]{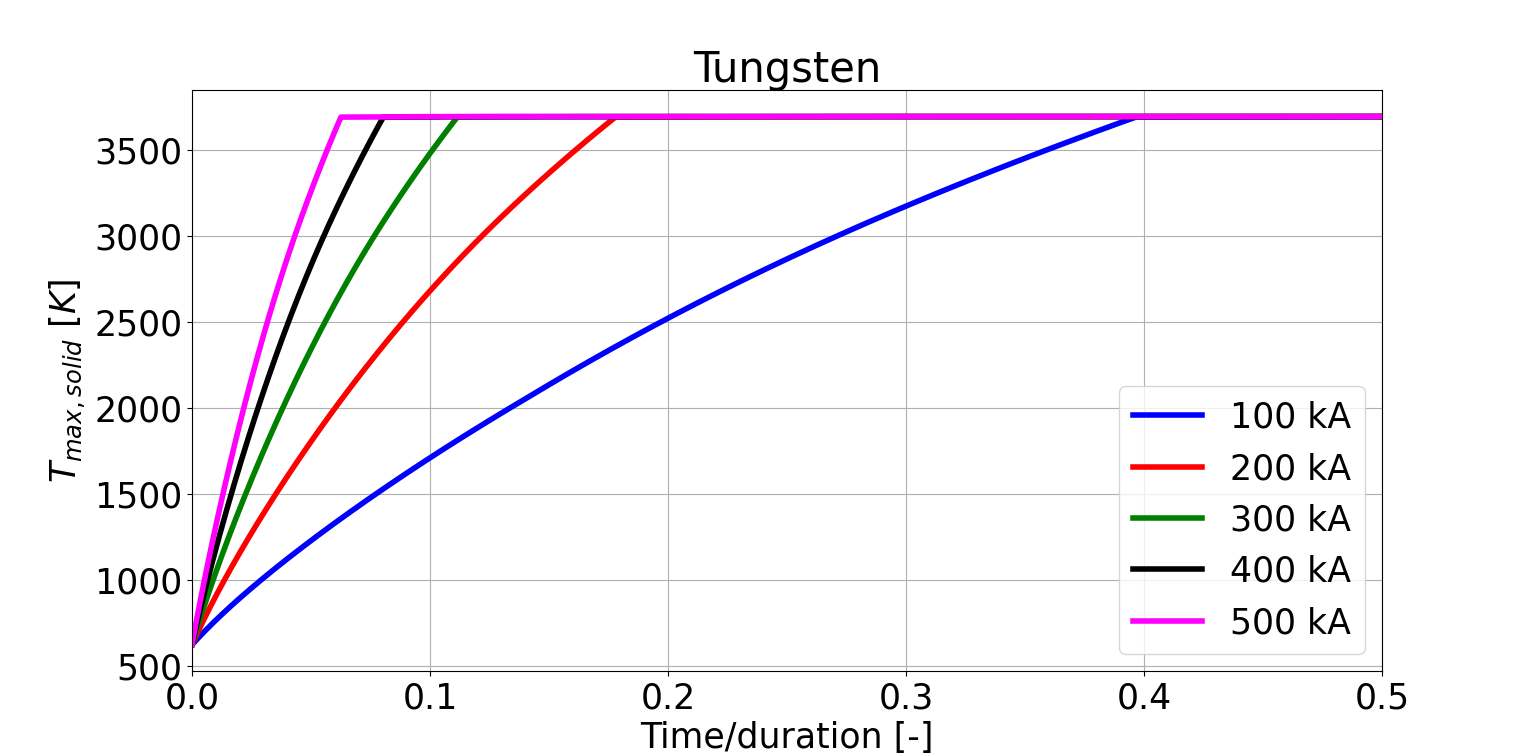}}
    \caption{Maximum solid temperature for (a) Beryllium and (b)  Tungsten during the discharge time.}
    \label{Tmaxsolid}
\end{figure}

The maximum solid temperature has been examined to monitor the thermal transient evolution, with its evolution implicitly indicating the commencement of melting, if applicable. Figures \ref{Tmaxsolid} (a) and \ref{Tmaxsolid} (b) depict the evolution of maximum solid temperature for Tungsten and Beryllium tiles, respectively.

Observably, Tungsten exhibits higher resistance to beam exposure, reaching melting temperature later with respect to Beryllium. This contrasting behaviour is attributed to the material properties and the distinct interactions between the material and the electrons.

Furthermore, utilizing equation \ref{MeltE}, evaluating the energy required to heat a unit volume of material from $T_0$ to $T_m$ and achieve complete melting is feasible. The analysis reveals that Tungsten necessitates nearly double the energy of Beryllium to complete this process. The energy per cubic meter needed to raise the material to the melting temperature and complete the melting process is documented in table \ref{EnergyMelting}.

\begin{equation}
    E_m = \rho \cdot \left ( \Delta h_f + \int_{T_0}^{T_m} c_p (T) dT \right )
    \label{MeltE}
\end{equation}

\begin{table}[!htbp]
    \centering
    \begin{tabular}{|c|c|c|c|}
        \hline
        \textbf{Material} & \textbf{$\rho \cdot \int_{T_0}^{T_m} c_p (T) dT \,\, \left [ \frac{GJ}{m^3} \right ]$} & \textbf{$\rho \cdot \Delta h_f \,\, \left [ \frac{GJ}{m^3} \right ]$} & \textbf{$E_{m} \,\, \left [ \frac{GJ}{m^3} \right ]$} \\
        \hline
        Tungsten & 5.24 & 11.2 & 16.4 \\
        \hline
        Beryllium & 2.96 & 5.24 & 8.20 \\
        \hline
    \end{tabular}
    \caption{energy per unit volume to melt Tungsten and Beryllium.}
    \label{EnergyMelting}
\end{table}

Considering both the greater energy deposition (and lower stopping power) and the reduced energy per unit volume needed to heat from the initial temperature and melt the Beryllium, the expected damage is greater in Beryllium than in Tungsten if the same RE impacts the tile.

\begin{figure}[!htbp]
    \centering
    \includegraphics[width = 0.95\linewidth]{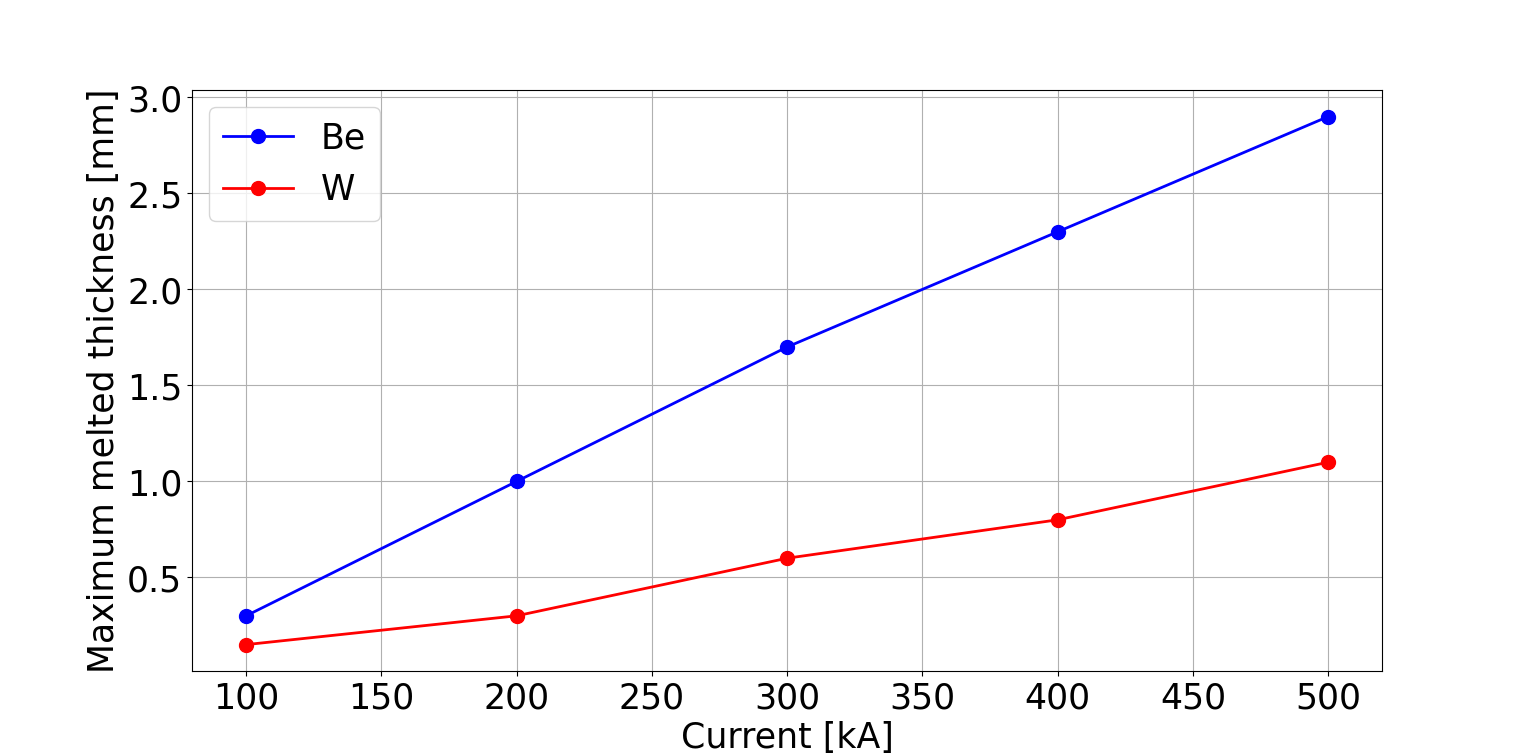}
    \caption{Melted thickness as a function of the beam RE current for the two considered tile materials.}
    \label{MeltThick}
\end{figure}

\begin{figure}[!htbp]
    \centering
    \includegraphics[width = \linewidth]{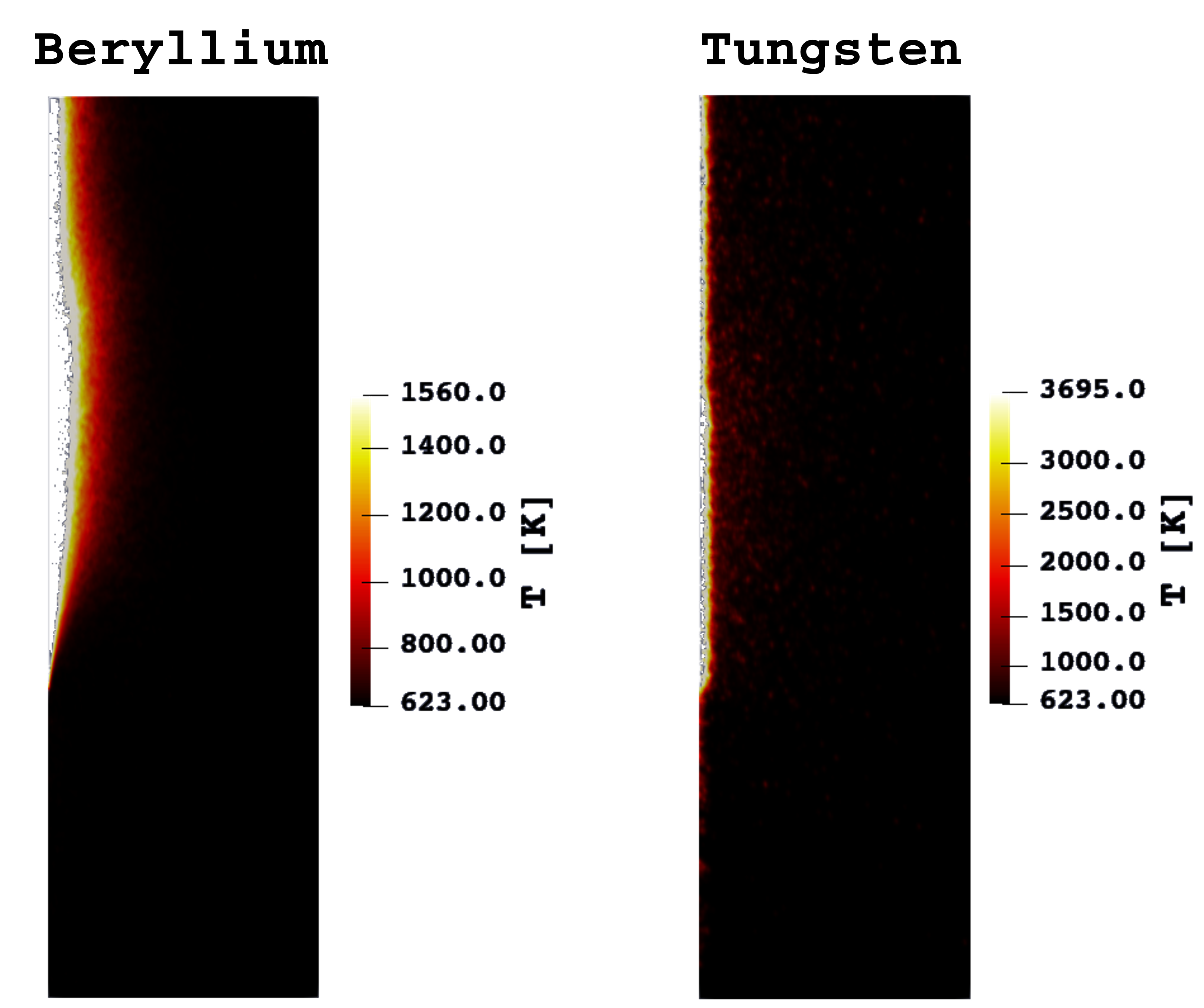}
    \caption{Temperature map of the entire solid region at the end of the energy deposition for a beam current of 500 kA. Beryllium at left and Tungsten at right. The original shape of the solid region is highlighted by a light grey line.}
    \label{Tmap500kA}
\end{figure}

The assessment of tile damage relies on measuring melted thickness at 7 cm from the bottom of the tile, where the maximum damage is detected due to the inclination of the beam, as depicted in Fig. \ref{MeltThick}, concerning the beam RE current for both materials. This allows confirmation of the anticipated greater damage in Beryllium. The discrepancy in damage is even more evident in the temperature map displayed in Fig. \ref{Tmap500kA}, where the molten portion of the geometry has been omitted to reveal the location of the melting front after energy deposition for a beam current of 500 kA. From the temperature map, it is evident how, in the case of tungsten, not only the melting front is formed later with respect to beryllium, but it also propagates much less in the thickness of the tile.

\section{\label{sec:discussion} Discussion}

The volumetric energy density profiles result from three electron transport mechanisms within the energy range of interest: Bremsstrahlung, elastic scattering (both scaling as $Z^2$), and delta ray production (scaling as Z). When focusing on the near-surface region, the contribution of photons generated through Bremsstrahlung to deposited energy is negligible, while elastic scattering and delta-ray production play a significant role. Notably, elastic scattering and delta ray production are more efficient in W than in Be, with the former characterized by a steep VED gradient localized in the near-surface region. Furthermore, at a higher RE impact angle, the energy deposited by the beam increases in both materials, enabling the beam to penetrate deeper into the material while conserving a significant amount of its initial energy.

The presence of the magnetic field significantly affects the VED profiles. This is due to the re-deposition of the back-scattered electrons energy near the initial impact point. Considering the case with 5° impact angle, in tab. \ref{tab:energy_magnetic}, the values of the total energy deposited in Be and W is reported in the presence (E w B) and absence (E w\/o B) of the magnetic field. The effects of the magnetic field are more remarkable in Tungsten with respect to Beryllium since the former is characterized by a more significant number of backscattered electrons whose redeposition causes an increase of 100\% of the energy deposited in the tile. Due to this reason, in Fig. \ref{fig:2D_BeW}, the near-surface region of the Tungsten tile at X>5 is characterized by a broadening of the high-intensity energy density region. The same occurs also in Be where, however, the effect is less evident in terms of the total energy deposited in the material due to a larger number of electrons penetrating into the material.

\begin{table}[!hbtp]
    \centering
    \begin{tabular}{|c|c|c|}
         \hline
        Material & E w B [MeV] & E w/o B [MeV] \\
         \hline
        Be & 5.17  & 4.57 \\
         \hline
        W & 5.23 & 2.80 \\
         \hline
        
    \end{tabular}
    \caption{Average energy deposited in Beryllium and Tungsten with an impact angle of 5° with (w B) and without (w/o B) magnetic field per primary.}
    \label{tab:energy_magnetic}
\end{table}

The broadening of the high intensity energy density region does not extinguish in the analyzed tile (still present at X=10 in Fig. \ref{fig:2D_BeW}). That is, the backscattered electrons re-deposition do not end in this tile. Therefore, the remaining energy not deposited in this tile would be deposited in an adjacent tile. If the 5° impact angle case is considered, the backscattered electrons will deposit around 2 MeV/primary for both materials in the adjacent tile.  

To evaluate the potential damage to the PFC resulting from material melting caused by the energy deposition from RE beams, a thermal model based on an apparent heat capacity formulation has been developed using FreeFEM++ and integrated into a Python wrapper. This model offers the flexibility to automatically process FLUKA output files and generate inputs for FreeFEM++, accommodating various geometries and loads.

The thermal model has undergone successful benchmarking against references in the literature and has subsequently been applied to a preliminary test case. This case involves the impact of a 10 $cm^2$ square RE beam on Tungsten and Beryllium tiles. Various RE beam currents from 100 kA to 500 kA were considered in the analysis, enabling the estimation of the expected melted thickness as a function of the current. 
In the considered range of currents, the W tile is characterized by a lower melted thickness due to W higher energy requirement per unit volume for reaching its melting temperature and completing the melting phase, as well as the beam-reduced penetration into the tile thickness.

\begin{figure}[!htbp]
    \centering
    \includegraphics[width = \linewidth]{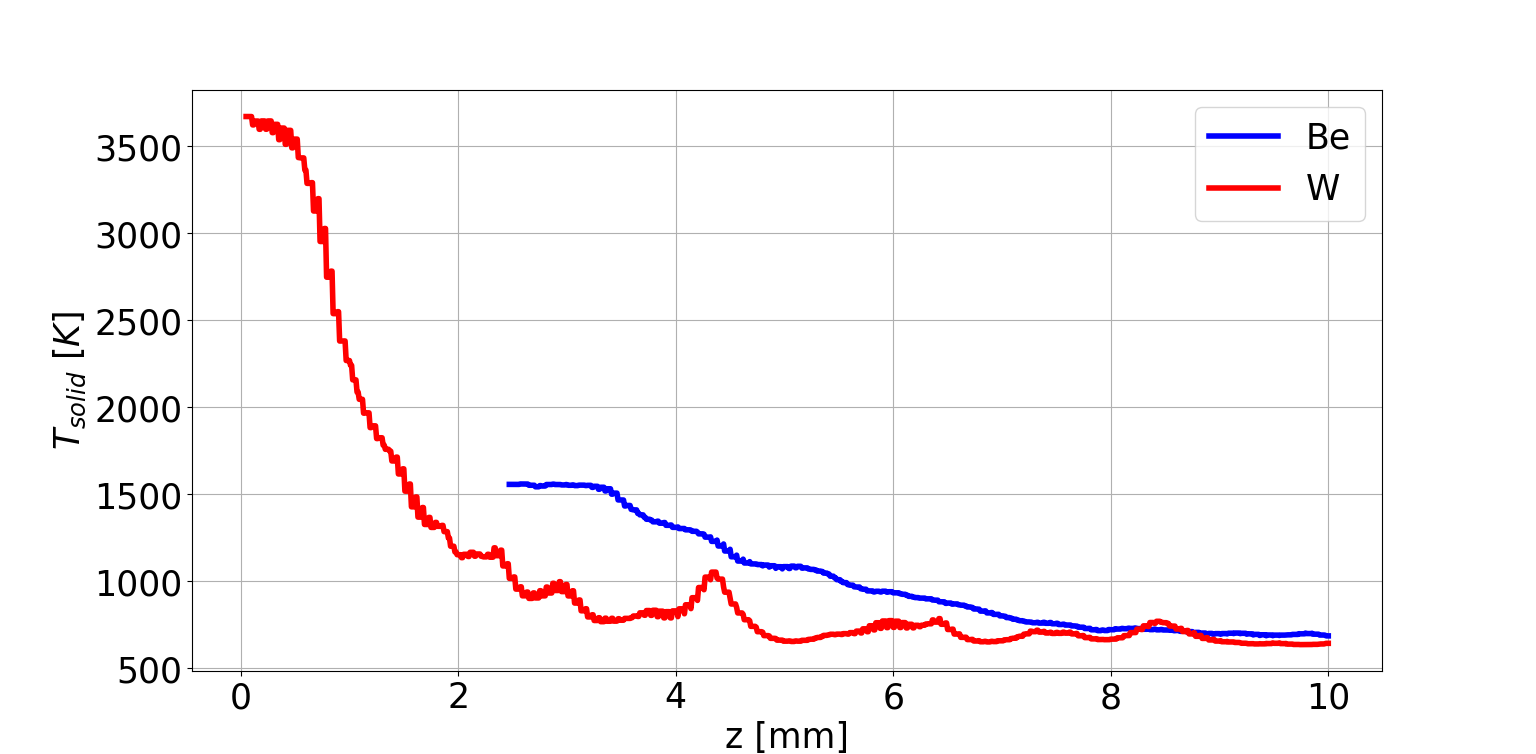}
    \caption{Temperature profile within the solid domain along a line at 7 cm from the tile bottom for both Tungsten and Beryllium in case of a 500 kA beam current. Only the first 10 mm of the total thickness are shown here, where relevant gradients can be observed}
    \label{Tprof}
\end{figure}

The analysis carried out with FreeFem++ make use of a 2D approximation. The temperature map shown in Fig. \ref{Tmap500kA} can be considered to justify this approximation. What should be compared here is the beam characteristic loss time and the characteristic time of the heat diffusion (computed as the ratio of the square of the characteristic dimension and the heat diffusivity). This characteristic time provides an estimate of the time required by the heat to diffuse in a characteristic length which is in our case the tail thickness. It is evident that the back of the tile, where power deposition is minimal, remains at the initial temperature due to insufficient time for heat diffusion. Furthermore, the temperature profile along the horizontal line passing through the beam centre (Fig. \ref{Tprof}) demonstrates that within 1 cm of tile thickness, the temperature remains unchanged from the initial temperature, confirming negligible heat diffusion in the considered  time interval. A comparison of the temperature profiles in the two materials reveals the qualitative difference in energy deposition effects. Despite the high temperature near the surface in tungsten, beryllium exhibits a smoother temperature gradient, resulting in higher temperatures beyond the initial millimeters of thickness compared to tungsten.

It is well known that the interaction between a RE beam and a material extends far beyond the initial melting simulated in this study.  For example, in Tungsten, the shorter penetration depth compared to Beryllium leads to an extremely high energy density near the surface, even at relatively low RE currents. This can trigger complex dynamics not simulated in this study.

Moroever, the transient does not stop at the end of the disruption and the heat diffusion makes the temperature at deeper thickness increase afterwards, potentially melting the cooling substrate of the first wall. However, in the analyzed condition, this is not expected. Indeed, estimating the final average tile temperature with a 0D model (which thus neglects heat diffusion) reported in equation \ref{0Dmodel}, it is possible to realize that this value is not worrying (< 920 K) and thus damage on the cooling zone is not foreseen.

\begin{equation}
    \label{0Dmodel}
    A \cdot \rho \cdot c_p \cdot (T-T_0)=\int_A E'''(x,y) dxdy
\end{equation}

Another aspect to take into consideration is the material resolidification, which is not taken into account in this model. However, in some sense, the 0D model reported in equation \ref{0Dmodel} accounts for the net balance of the specific heats. It neglects the subtraction of heat required for the material melting. Still, at the same time, it neglects the positive heat given back by the material once it solidifies, and the obtained temperature is far from being dangerous for the other possible materials interfaced with the PFC.

\section{\label{sec:conclusion} Conclusions}

The work discusses the assessment of damage caused by a 7 MeV RE beam in fusion reactor FW materials such as Beryllium and Tungsten using FLUKA and FreeFem++ simulation tools. 

FLUKA was utilized to calculate the energy deposition profiles resulting from the interaction of an RE beam with a sample tile made of Be and W. On the other hand, the apparent heat capacity formulation was implemented and benchmarked using FreeFem++. Once validated, the model was used to compute the temperature and melting front evolution in the materials using the energy deposition profiles from FLUKA as an input.

Realistic RE energy distribution functions derived from the observation of a RE beam in the \#39012 experiment of ASDEX Upgrade  were considered. Two different beam sizes were analyzed for Be and W materials in the presence of a magnetic field at the wall. The study assumed a point beam to investigate the role of three transport mechanisms (elastic scattering, delta ray production, and Bremsstrahlung) with an impact angle of 5 degrees. W, with its high stopping power due to its high atomic number, experiences a steep gradient of the volumetric energy density deposited by RE near the surface, where elastic scattering and delta-ray production are dominant mechanisms. On the other hand, the dynamics of photon transport were investigated by turning off Bremsstrahlung. 

In Be, energy deposited by photons becomes dominant near the opposite border of the tile, while in W, after a few millimetres, photons become the main energy carrier. This switch between electrons and photons as the main energy carrier significantly changes the energy deposition pattern, as photons, with their high radiation length, can carry their energy deeper into the material. Therefore, even though the deposited energy decreases rapidly near the surface in W, far from the surface, the deposited energy is higher in W than Be in terms of MeV/cm per primary electron. The penetration of photons has no consequence for W due to its good thermal properties, but it may affect other structural materials (like in the Breeding Blanket) after the initial W protection. 

In addition, a higher impact angle enables the beam to penetrate deeper into the material while conserving a substantial amount of its initial energy. As a consequence, the generation of photons becomes even more problematic for the FW materials. 

For thermal analysis, a 10x10 cm square RE beam with a 5-degree impact angle was considered, which confirmed W capability to limit the melting front to the near-surface region where a significant amount of energy is deposited. Moreover, a higher RE current is needed for W than for Be to initiate tile melting, due to W higher energy required to heat a unit volume to the melting temperature. However, in reactor scale devices higher runaway currents and runaway energies than those considered here are possible. In addition, such steep gradients in W can initiate complex phenomena such as vaporization or material explosion which are not modeled in this thermal analysis.

However, the primary objective of this work is to estimate the melted thickness of the PFC, for which the current model is sufficient. Indeed, at the end of the RE impact, heat begins to diffuse from the melting front—the hottest point in the solid region—causing the rest of the tile to heat up. However, due to the material homogeneity, its temperature will never exceed the melting point. As a result, simulating only the beam loss duration provides a reasonable estimate of the melted thickness. Future work will focus on analyzing more complex dynamics by incorporating additional details into the model.

\section{\label{sec:level10}Acknowledgements}
We thank Dr Ondrej Ficker and Dr Andrea Dal Molin for providing the input data used for FLUKA simulations and Dr Francesc Salvat Pujol for the fruitful discussions that helped us understand the behaviour of the electrons in the material. In addition, we thank the reviewers whose valuable comments allowed us to improve this manuscript.
This work has been carried out within the framework of the EUROfusion Consortium, funded by the European Union via the Euratom Research and Training Programme (Grant Agreement No 101052200 — EUROfusion). Views and opinions expressed are however those of the author(s) only and do not necessarily reflect those of the European Union or the European Commission. Neither the European Union nor the European Commission can be held responsible for them.

\section{\label{sec:level10}Data Availability}
The data that support the findings of this study are available from
 the corresponding author upon reasonable request.

 \section{Appendix}
 \label{benchmark}

 The reference problem involves heating the front face of an infinite slab with a constant heat flux, initiating a melting front. The transient melting process is monitored through the temperature measurements of the front and back faces, along with the position of the melting front. For details regarding boundary conditions and material properties, refer to \cite{zerroukat}.

\begin{figure}[!htbp]
    \centering
    \includegraphics[scale=0.35]{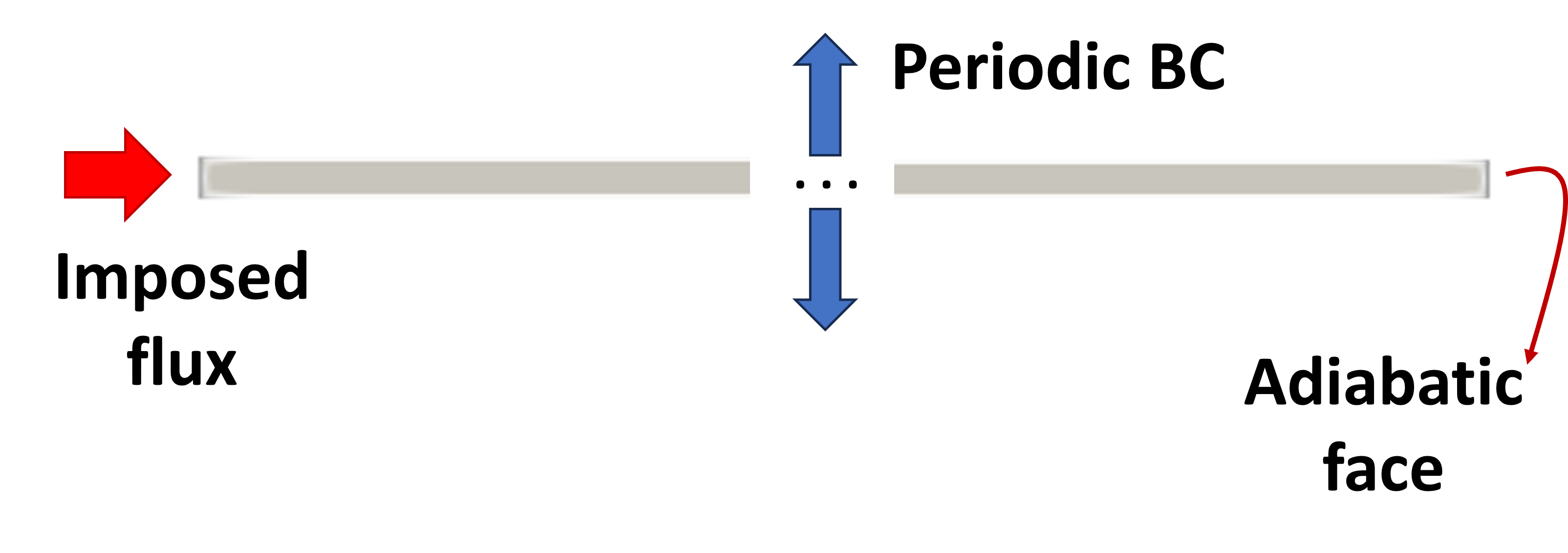}
    \caption{Computational domain for solving the benchmark problem and summary of the adopted BCs. The slab width dimension has been cut to improve readability.}
    \label{domain}
\end{figure}

Although the original problem is inherently 1D, to assess the functionality of the 2D FreeFEM model, the infinite slab has been represented in 2D by a thin slice with the same length as the width of the slab. Furthermore, periodic boundary conditions have been imposed on the upper and lower faces. Fig. \ref{domain} depicts the computational domain and boundary conditions.

\begin{figure}[!htbp]
    \centering
    \includegraphics[scale = 0.3]{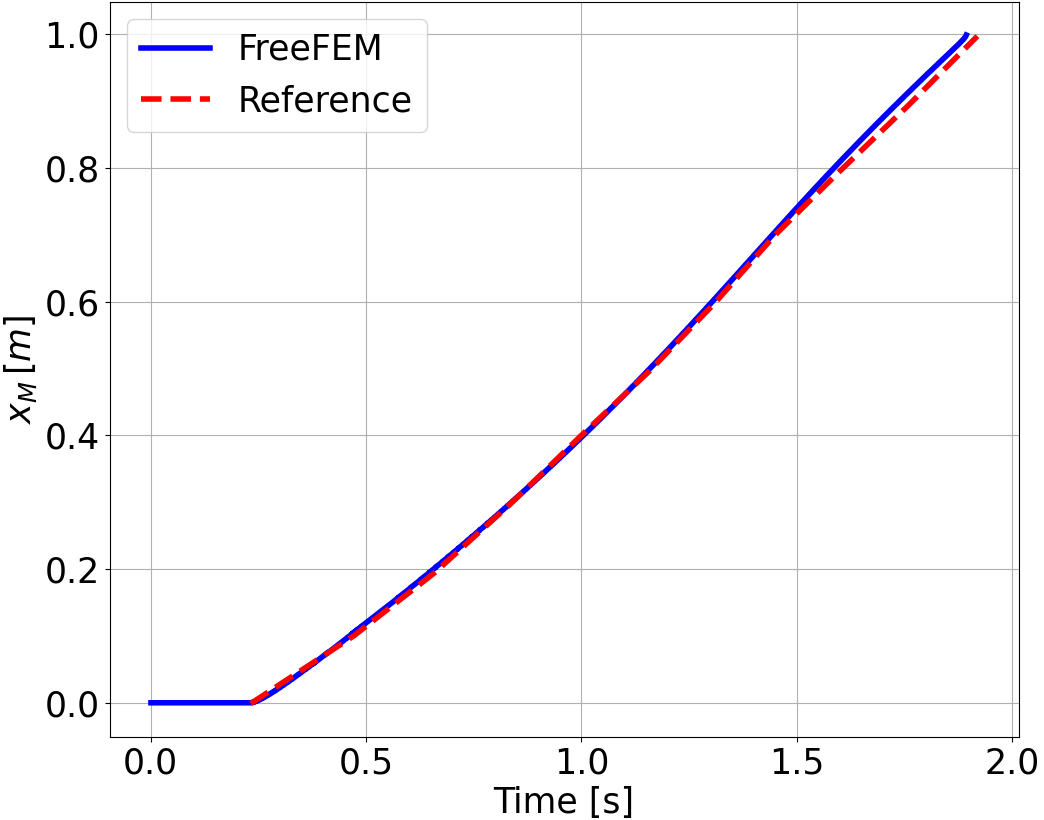}
    \caption{Position of the melting front in the benchmark problem. Comparison between the results obtained with the apparent heat capacity method developed in FreeFEM versus the reference results \cite{zerroukat}.}
    \label{xM}
\end{figure}

\begin{figure}[!htbp]
    \centering
    \includegraphics[scale = 0.3]{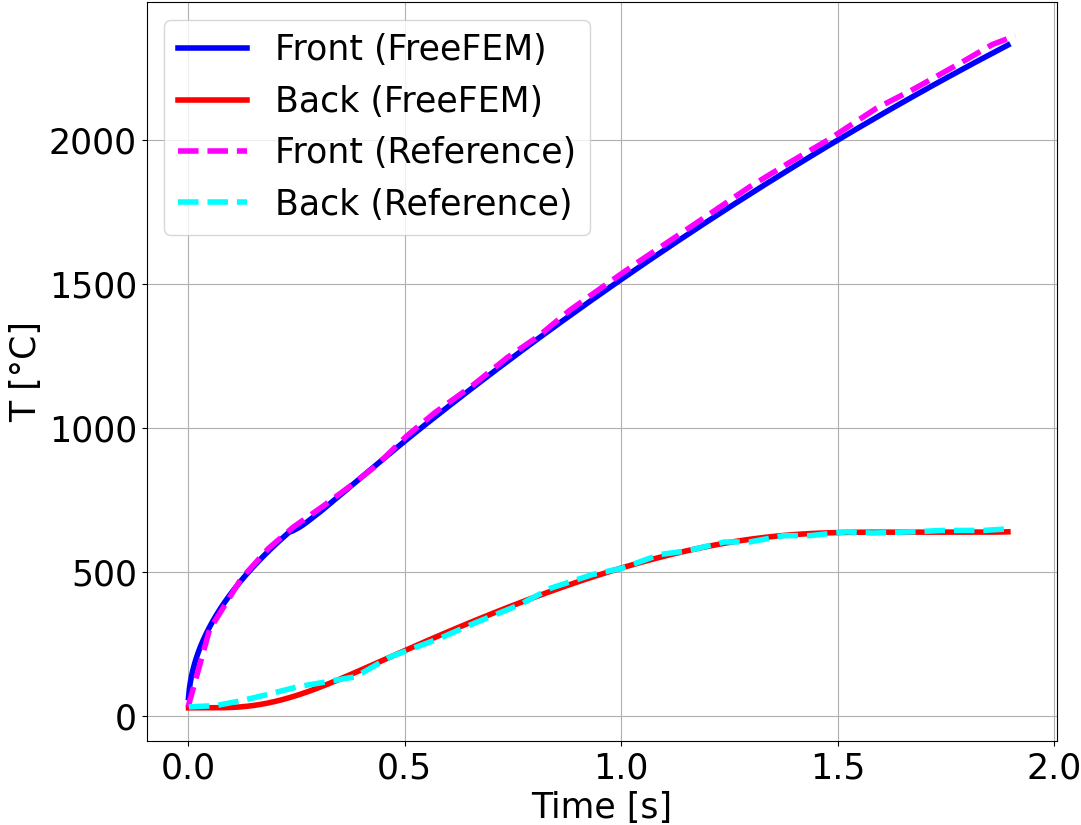}
    \caption{Front and back face temperature in the benchmark problem. Comparison between the results obtained with the apparent heat capacity method developed in FreeFEM versus the reference results \cite{zerroukat}.}
    \label{Temp1b}
\end{figure}

In the present study, the results obtained from the finite element model have been compared with the findings reported by \cite{zerroukat} and depicted in figures \ref{xM} and \ref{Temp1b}. The comparison revealed a high level of concurrence in terms of the local temperature evolution and the position of the melting front. This underscores the model proficiency in accurately replicating the physics of the specified reference scenario.

\newpage
\nocite{*}
\bibliography{aipsamp}

\providecommand{\noopsort}[1]{}\providecommand{\singleletter}[1]{#1}
\begin{thebibliography}{10}

\bibitem{Bandaru2025}
V.~Bandaru, M.~Hoelzl, F.J. Artola, M.~Lehnen, and et~al.
\newblock {\em J. Plasma Phys.}, 91(1), 2025.

\bibitem{Vannini2025}
F.~Vannini, V.~Bandaru, H.~Bergstrom, N.~Schwarz1, and et~al.
\newblock {\em Nucl. Fusion}, Accepted manuscript, 2025.

\bibitem{Bartels1993}
H.-W. Bartels.
\newblock {\em Fusion Eng. Des.}, 23:323--328, 1993.

\bibitem{Igitkhanov2011b}
Yu. Igitkhanov, B.~Bazylev, and I.~Landman.
\newblock {\em J. Nucl. Mater.}, 415:S845--S848, 2011.

\bibitem{Caloud2024}
J.~Caloud, E.~Tomesova, O.~Ficker, J.~Cerovsky, and et~al.
\newblock {\em Rev. Sci. Instrum.}, 95:113512, 2024.

\bibitem{Ratynskaia2025}
S.~Ratynskaia, P.~Tolias, T.~Rizzi1, K.~Paschalidis, and et~al.
\newblock {\em Nucl. Fusion}, 65:024002, 2025.

\bibitem{Bazylev2011}
B.~Bazylev, G.~Arnoux, W.~Fundamenski, Yu. Igitkhanov, and et~al.
\newblock {\em J. Nucl. Mater.}, 415 (1):S841--S844, 2011.

\bibitem{Bazylev2007}
B.~Bazylev, G.~Janeschitz, I.~Landman, S.~Pestchanyi, and et~al.
\newblock {\em Phys. Scr.}, T128:229--233, 2007.

\bibitem{Ahdida2022}
C.~Ahdida, D.~Bozzato, D.~Calzolari, F.~Cerutti, and et~al.
\newblock {\em Front. Phys.}, 9:788253, 2022.

\bibitem{Battistoni2015}
G.~Battistoni, T.~Boehlen, F.~Cerutti, P.W. Chin, and et~al.
\newblock {\em Ann. Nucl. Energy}, 82:10--18, 2015.

\bibitem{COMSOL}
Comsol ab, comsol multiphysics® v. 5.6.
\newblock \url{www.comsol.com}.
\newblock Accessed: 2025-02-09.

\bibitem{Carbajal2017}
L.~Carbajal, D.~del Castillo-Negrete, D.~Spong, S.~Seal, and et~al.
\newblock {\em Phys. Plasmas}, 24:042512, 2017.

\bibitem{Allison2016}
J.~Allison, K.~Amako, J.~Apostolakis, P.~Arce, and et~al.
\newblock {\em Nucl. Instrum. Methods Phys. Res. A}, 835:186--225, 2016.

\bibitem{MR3043640}
F.~Hecht.
\newblock {\em J. Numer. Math.}, 20(3-4):251--265, 2012.

\bibitem{ComsolDOC}
{\em COMSOL Documentation. Version 6.1 - Heat Transfer Module User's Guide}.

\bibitem{richiusa2023advances}
M.L. Richiusa, P.~Ireland, F.~Maviglia, J.~Nicholas, and et~al.
\newblock {\em Fusion Eng. Des.}, 189:113477, 2023.

\bibitem{Ferrari1992}
A.~Ferrari, P.R. Sala, , R.~Guaraldi, and F.~Padoani.
\newblock {\em Nucl. Instrum. Methods Phys. Res. B}, 71 (4):412--426, 1992.

\bibitem{Fer05}
A.~Ferrari, J.~Ranft, P.R. Sala, and A.~Fassò.
\newblock {\em FLUKA: A multi-particle transport code (Program version 2005)}.

\bibitem{Kim86}
L.~Kim, R.H. Pratt, S.M. Seltzer, and M.J. Berger.
\newblock {\em Phys. Rev. A}, 33, 1986.

\bibitem{NIST}
Stopping powers and range tables for electrons.
\newblock \url{https://physics.nist.gov/PhysRefData/Star/Text/ESTAR.html}.
\newblock Accessed: 2023-07-27.

\bibitem{zerroukat}
M.~Zerroukat and C.R. Chatwin.
\newblock {\em Computational Moving Boundary Problems. Chapter 6, page 163}.
\newblock Research Studies Press LTD, Taunton, Somerset, England, 1994.

\bibitem{gupta1981variable}
R.S. Gupta and D.~Kumar.
\newblock {\em Int. J. Heat Mass Transf.}, 24(2):251--259, 1981.

\bibitem{caldwell2003nodal}
J.~Caldwell, S.~Savovic, and Y.Y. Kwan.
\newblock {\em J. Heat Transf.}, 125(3):523--527, 2003.

\bibitem{caldwell2004numerical}
J.~Caldwell and Y.Y. Kwan.
\newblock {\em Commun. Numer. Meth. En.}, 20(7):535--545, 2004.

\bibitem{Eurofusion_Wiki}
Wpte wikipages: Experimental campaign 2021:rt05.
\newblock \url{https://wiki.euro-fusion.org/wiki/WPTE\_wikipages:\_Experimental\_campaign\_2021:RT05}.
\newblock Accessed: 2025-01-24.

\bibitem{DalMolin2023}
A.~Dal Molin, M.~Nocente, M.~Dalla Rosa, E~Panontin, and et~al.
\newblock {\em Meas. Sci. Tech.}, 34:085501, 2023.

\bibitem{Salvat2006}
F.~Salvat, J.M. Fernández-Varea, J.~Sempau, and X.~Llovet.
\newblock {\em Radiat. Phys. Chem.}, 75 (10):1201--1209, 2006.

\bibitem{TechReport_2022_LANL_LA-UR-22-30006Rev.1_KuleszaAdamsEtAl}
J.~A. Kulesza, T.~R. Adams, J.C. Armstrong, S.R. Bolding, and et~al.
\newblock {MCNP\textsuperscript{\textregistered} Code Version 6.3.0 Theory \& User Manual}.
\newblock Technical Report LA-UR-22-30006, Rev.~1, Los Alamos National Laboratory, Los Alamos, NM, USA, September 2022.

\bibitem{DalMolin2020}
A.~Dal Molin.
\newblock PhD thesis, University of Milano-Bicocca, 2020.

\bibitem{Ondrej}
Ondrej Ficker.
\newblock Private comunication.
\newblock 2024.

\bibitem{tolias2017W}
P.~Tolias, EUROfusion~MST1 Team, and et~al.
\newblock {\em Nucl. Mater. Energy}, 13:42--57, 2017.

\bibitem{tolias2022Be}
P.~Tolias.
\newblock {\em Nucl. Mater. and Energy}, 31:101195, 2022.

\end{thebibliography}
\end{document}